\documentclass[preprint,prl,10pt]{revtex4}%
\usepackage{amsfonts}
\usepackage{amsmath}
\usepackage{amssymb}
\usepackage{graphicx}%
\begin{document}
\preprint{cond-mat}
\title[ ]{Correlation Properties of the Electron-Hole Plasma Interecting with the
Exciton Gas and the Formation of Inhomogeneous State. }
\author{V. S. Babichenko}
\affiliation{R.S.C. "Kurchatov Institute" }
\author{}
\affiliation{}
\author{}
\affiliation{}
\pacs{PACS number}

\begin{abstract}
The correlation properties of the cold system consisting of the electron-hole
plasma interacting with the exciton gas are analyzed. It is shown that the
homogeneous state of the system is unstable and in the stationary state the
densities of the electron-hole plasma and exciton gas are modulated.

\end{abstract}
\volumeyear{ }
\volumenumber{ }
\issuenumber{ }
\eid{ }
\date{}
\startpage{1}
\endpage{ }
\maketitle

The problem of the existence of the Bose condensed state of the exciton gas in
semiconductors is the subject of the experimental and theoretical
investigations for a long time. However, there is no undoubted proof of the
existence of this state until now.

Recently, the properties of the cold exciton gas are investigated intensively,
both experimentally and theoretically, in double 2D quantum wells, besides,
electrons are localized on one plane of these wells and the holes are
localized on the other plane. The state of the cold exciton system being
observed in the experiments of different experimental groups \cite{B1},
\cite{Sn1}, \cite{Sn2}, \cite{Tim}, \cite{B2}, \cite{Sn3}, \cite{Sn4},
\cite{Tim1}, \cite{Tim2}, \cite{Tim3} is nonhomogeneous, and the density of
the system in these experiments is modulated in space.

The theory of this phenomenon based on the Turing kinetic mechanism of the
instability \cite{Tur} have been proposed in works \cite{B2}, \ \cite{B3}. In
this theory the system of nonlinear diffusion equations with sources of
particles for electrons, holes and excitons are proposed and solved. The
solution of these equations demonstrates the state with the periodically
modulated density of the exciton system only if the constant describing the
decay of excitons depends on their density. However, in the low density
approximation this dependence should be neglected. Moreover, this mechanism
has the classical character and does not describe the systems with the quantum
coherency, but for the sufficiently small temperatures one can suppose the
existence of the coherency in the exciton system \cite{K1}, \cite{K2},
\cite{K4}.

Another theory, based on the supposition of the existence of the attraction
between excitons at large distances and the formation of the liquid phase in
the low-density exciton system due to this attraction, has been proposed in
the work \cite{Sug}. The supposition of the existence of the liquid phase in
the low-density exciton system is baseless. The existence of the attractive
part of the interparticle interaction potential in 2D Bose systems results in
the collapse of the system to the densities of the order of the squared
inverse radius of the repulsive core \cite{Bab}. However, for these densities
excitons can not exist as individual particles, they are destroyed and form
the electron-hole liquid. Note that in the usual 3D case, as it was shown in
the work \cite{K1}, the scattering amplitude for the isotropic excitons is
positive and is equal to the value $a=\frac{13\pi}{3}a_{B}$, where $a_{B}$ is
the effective Bohr radius, and just the scattering amplitude, but not the bare
potential, determines the properties of the system in the case of small
densities. The effective Bohr radius has the form $a_{B}=\frac{\hbar^{2}%
}{m^{\ast}\left(  e^{\ast}\right)  ^{2}}$; where $\hbar$ is the Planck
constant; $m^{\ast}=\frac{m_{e}m_{h}}{m_{e}+m_{h}}$ is the reduced mass,
$m_{e,h}$ are electron (e) and hole (h) masses, which are supposed to be
isotropic and of the same order $m_{e}\sim m_{h}$; $e^{\ast}$ is the effective
charge in the semiconductor $e^{\ast}=\left(  \frac{e^{2}}{\varkappa_{0}%
}\right)  ^{1/2}$; where $e$ is the bare electron charge and $\varkappa_{0}$
is the static dielectric constant of the semiconductor. As the result, the 3D
exciton system of small density can exist only in the gas phase. The repulsion
for the interaction between excitons in 2D system, where electrons and holes
are localized on different planes, is sharply defined than in the usual 3D
case due to the geometrical constraint. In this case the formation of the
exciton liquid phase is possible only for the densities $n_{ex}\geq
1/a_{ex}^{2}$, where $a_{ex}$ is the size of the exciton. The size of the
exciton is equal to the effective Bohr radius $a_{ex}=a_{B}$ in the case of
the small distance $l\ $between the planes of the quantum wells $l<<a_{B}$,
and $a_{ex}>>a_{B}$ for the large distance between planes when $l>>a_{B}$. If
the exciton density is sufficiently large $n_{ex}\geq1/a_{ex}^{2}$ the
excitons can not be considered as the individual particles. In this case the
electron-hole system represents the strongly interacting electron-hole liquid.

In the present work the Coulomb correlations in the cold electron-hole plasma
as well as the quantum effects and the coherency of the exciton subsystem in
quantum wells are taken into account. We suppose that the external source is
stationary, has the frequency larger than the semiconductor gap $\Delta_{g}$
and creates electrons and holes being localized on different planes. Moreover,
the region of the action of the electromagnetic field is supposed to be
sufficiently small. In this case the electron-hole plasma being created by the
external source propagates along 2D planes, besides, the electrons propagate
along one plane and the holes along the other plane. During this propagation
the electron-hole plasma loses the energy giving away the energy to the phonon
system and comes to the equilibrium state with the small temperature T. Note
that the parameters of the stationary state, such that the densities of the
electron-hole plasma $n$ and the exciton gas $n_{ex}$, are defined by the
equilibrium between incoming rate of the electrons, holes and excitons and
outgoing rate of these particles.

Thus, the model being analyzed represents the quasi-equilibrium electron-hole
plasma from which the exciton gas is created as the result of the formation of
the electron-hole bound states. In the present work we consider the properties
of the stationary state of this system, namely, the properties of the
electron-hole plasma of the small density $na_{B}^{3}<<1$ interacting with the
exciton gas of small density $n_{ex}a_{B}^{3}<<1$.

The small density of the electron-hole plasma and the exciton gas are
necessary if we suppose the existence both electron-hole plasma and the
exciton gas. In the opposite case of the large density $na_{B}^{3}>>1$ the
excitons are destroyed and can not exist as the individual particles. This
case is not under consideration in this work. Note that the experiments
\cite{B1}, \cite{Tim}, \cite{B2} demonstrate the existence of the excitons,
and this means the existence of small density of the electron-hole plasma in
these experiments.

We analyze the correlation properties of the electron-hole plasma of the small
density and the interaction of this plasma with the exciton gas. The energy of
the ground state of the electron-hole plasma has a minimum value as the
function of the density at some density $n_{0}$ such that $n_{0}a_{B}^{2}%
\sim1$. As a result, the electron-hole plasma tends to form the liquid
electron-hole drops and the homogeneous state of the system becomes unstable.
The inhomogeneous state of the electron-hole plasma creates the periodic mean
field which acts on the exciton gas and makes the density of the exciton
system modulated in space.

We suppose that the system consisting of the electron-hole plasma interacting
with the exciton gas is in the stationary state with some small temperature T
obeying the inequality $T<<\varepsilon_{F}^{\left(  e,h\right)  }%
<<E_{b}^{\left(  ex\right)  }$; where $\varepsilon_{F}^{\left(  e,h\right)  }$
are the Fermi energies of the electrons and the holes correspondingly
$\varepsilon_{F}^{\left(  e,h\right)  }=\frac{p_{F}^{2}}{2m_{e,h}}$; the Fermi
momentum $p_{F}$ in 2D case is expressed by the equality $p_{F}=\left(  2\pi
n\right)  ^{1/2}$; the average density $n$ of the electron-hole plasma is
equal to the density of each component $n=n_{e}=n_{h}$; $E_{b}^{\left(
ex\right)  }$ is the exciton bound energy $E_{b}^{\left(  ex\right)  }%
=\frac{\hbar^{2}}{2m^{\ast}a_{B}^{2}}$. Below we use the system of units in
which $\hbar=m^{\ast}=e^{\ast}=1$.

\section{Effective action}

To analyze the properties of the exciton system interacting with the
electron-hole plasma the effective action for the slow fields is derived. In
this paper the properties of the stationary state of this system is under
consideration, but in future the kinetics of the formation of this state is
supposed to be analyzed. For this reason the Keldysh-Schwinger technique for
the nonequilibrium processes \cite{K3}, \cite{Schw}\ in the functional
integral formulation (see \cite{BK}, for example) is used for the derivation
of the effective action of the slow fields. The generation functional for the
considering system can be written as%

\begin{equation}
Z=\int%
{\displaystyle\prod\limits_{\alpha}}
D\psi_{\alpha}D\overline{\psi}_{\alpha}\exp\left(  iS_{e-h}+i%
{\displaystyle\oint}
dtd^{2}r\left(  \overline{J}_{\alpha}\psi_{\alpha}+\overline{\psi}_{\alpha
}J_{\alpha}\right)  \right)  \label{Z}%
\end{equation}

The action of the electron-hole plasma in the electron-hole representation has
the form%

\begin{equation}
S_{eh}=%
{\displaystyle\oint}
d^{3}xd^{3}x^{\prime}\left\{
\begin{array}
[c]{c}%
{\displaystyle\sum\limits_{\alpha=e,h}}
\overline{\psi}_{\alpha}\left(  x\right)  \left[  i\partial_{t}+\mu_{\alpha
}-\varepsilon_{\alpha}\left(  \widehat{\overrightarrow{p}}\right)  \right]
\psi_{\alpha}\left(  x\right)  \delta\left(  x-x^{\prime}\right)  -\\
-\frac{1}{2}%
{\displaystyle\sum\limits_{\alpha,\beta}}
\left(  -1\right)  ^{\alpha+\beta}\overline{\psi}_{\alpha}\left(  x\right)
\psi_{\alpha}\left(  x\right)  \widehat{U}_{C}\overline{\psi}_{\beta}\left(
x^{\prime}\right)  \psi_{\beta}\left(  x^{\prime}\right)  \delta\left(
t-t^{\prime}\right)
\end{array}
\right\}  \label{Seh}%
\end{equation}

here $x=\left(  t,\overrightarrow{r}\right)  $ denotes the collection of time
t and space coordinates $\overrightarrow{r}$\ in 2D; $\psi_{e}$ and $\psi_{h}$
are fermion (Grassmann) fields of electrons and holes; $\varepsilon_{e}\left(
\widehat{\overrightarrow{p}}\right)  $ and $\varepsilon_{h}\left(
\widehat{\overrightarrow{p}}\right)  $ are the dispersive laws of the electron
and hole bands $\varepsilon_{\alpha}\left(  \widehat{\overrightarrow{p}%
}\right)  =\frac{1}{2}\Delta_{g}+\frac{1}{2m_{\alpha}}\widehat{\overrightarrow
{p}}^{2}$, where $\Delta_{g}$ is the semiconductor gap; $\widehat
{\overrightarrow{p}}=-i\overrightarrow{\nabla}$ is the momentum operator;
$\widehat{U}_{C}=U_{C}\left(  \overrightarrow{r}-\overrightarrow{r}^{\prime
}\right)  =\frac{e^{2}}{\varepsilon_{0}\mid\overrightarrow{r}-\overrightarrow
{r}^{\prime}\mid}=\frac{1}{\mid\overrightarrow{r}-\overrightarrow{r}^{\prime
}\mid}$ is the Coulomb interaction. In the Keldysh-Schwinger technique the
time variable t changes along the double time contour with return.

The effective action for the electron-hole plasma with the formation of the
exciton system have been obtained in the functional-integral technique in
\cite{BKex}, \cite{BKex1}. In this paper we give more simple derivation of the
effective action and introduce some simplification transforming to the density
functional for the Fermi subsystem. Integrating over the rapid $\psi_{\alpha}%
$-fields, changing on the scale much smaller than the average distance between
the charge carriers $n^{-1/3}$, in the ladder approximation we can obtain the
action for the smooth fields in the form%

\begin{equation}
S_{eh}^{\left(  s\right)  }=%
{\displaystyle\oint}
d^{3}xd^{3}x^{\prime}\left\{
\begin{array}
[c]{c}%
{\displaystyle\sum\limits_{\alpha=e,h}}
\overline{\psi}_{\alpha}\left(  i\partial_{t}-\xi_{\alpha}\right)
\psi_{\alpha}\delta\left(  x-x^{\prime}\right)  -\\
-\frac{1}{2}\left(  \overline{\psi}_{e}\psi_{e}-\overline{\psi}_{h}\psi
_{h}\right)  \widehat{U}_{C}\left(  \overline{\psi}_{e}\psi_{e}-\overline
{\psi}_{h}\psi_{h}\right)  \delta\left(  t-t^{\prime}\right)  -\frac{1}{2}%
{\displaystyle\sum\limits_{\alpha,\beta=e,h}}
\overline{\psi}_{\alpha}\psi_{\alpha}\Gamma_{\alpha\beta}\overline{\psi
}_{\beta}\psi_{\beta}%
\end{array}
\right\}  \label{Sgam}%
\end{equation}

here the fields $\psi_{\alpha}$\ are the smooth Fermi fields. The vertex
$\Gamma_{\alpha\beta}$ is the sum of the ladder diagrams. The internal Green
functions of these diagrams have the large momentums, much larger than $p_{F}%
$, and the external lines of these diagrams have the small momentums, much
smaller than the internal lines. The value of the boundary momentum $\Lambda$
separating the rapid and the smooth fields obeys the inequality $\frac{\hbar
}{a_{B}}>>\Lambda>>p_{F}$. The vertex $\Gamma$ can be considered as
independent of the external momentums or frequencies due to the large values
of the momentums corresponding to the internal lines of this vertex, on the
assumption that this vertex does not contain the pole. The vertexes
$\Gamma_{ee}$ and $\Gamma_{hh}$ do not contain the pole parts, but
$\Gamma_{eh}$ has the pole part corresponding to the formation of the bound
state of the electron and the hole, i.e. the exciton,%

\begin{equation}
\Gamma_{eh}\left(  P,k,k^{\prime}\right)  =\left(  E-\frac{\overrightarrow
{k}^{2}}{2m^{\ast}}\right)  \int\frac{d^{2}\overrightarrow{q}}{\left(
2\pi\right)  ^{2}}\left\{  \left[
{\displaystyle\sum\limits_{n}}
\frac{\psi_{n}\left(  \overrightarrow{k}\right)  \psi_{n}^{\ast}\left(
\overrightarrow{q}\right)  }{E-E_{ex}^{\left(  n\right)  }+i\delta}+\int
\frac{d^{2}\overrightarrow{p}}{\left(  2\pi\right)  ^{2}}\frac{\psi
_{\overrightarrow{p}}\left(  \overrightarrow{k}\right)  \psi_{\overrightarrow
{p}}^{\ast}\left(  \overrightarrow{q}\right)  }{E-E_{\overrightarrow{p}%
}+i\delta}\right]  U_{C}\left(  \overrightarrow{q}-\overrightarrow{k}^{\prime
}\right)  \right\}  \label{Ga}%
\end{equation}

The wave functions $\psi_{n}\left(  \overrightarrow{k}\right)  $ are the
relative motion wave functions of the discreet part of the spectrum of the
electron-hole pair, $\psi_{\overrightarrow{p}}\left(  \overrightarrow
{k}\right)  $ are the wave functions of the continuous part of the spectrum of
the electron-hole pair; the energies $E_{ex}^{\left(  n\right)  }$ are the
spectrum of the bound energies of the exciton; here we denote $E=\Omega
+\mu_{e-h}-\overrightarrow{P}^{2}/2M$, where $\mu_{e-h}=p_{F}^{2}/2m^{\ast}$.
The incoming $\overrightarrow{p}_{e}$, $\overrightarrow{p}_{h}$ and outgoing
$\overrightarrow{p}_{e}^{\prime}$, $\overrightarrow{p}_{h}^{\prime}$ momentums
of the ladder diagrams can be represented in the form
\begin{align*}
\overrightarrow{p}_{e}  &  =\frac{m_{e}}{M}\overrightarrow{P}+\overrightarrow
{k};\text{ \ \ \ \ \ \ }\overrightarrow{p}_{h}=\frac{m_{h}}{M}\overrightarrow
{P}-\overrightarrow{k}\\
\overrightarrow{p}_{e}^{\prime}  &  =\frac{m_{e}}{M}\overrightarrow
{P}+\overrightarrow{k}^{\prime};\text{ \ \ \ \ \ \ }\overrightarrow{p}%
_{h}^{\prime}=\frac{m_{h}}{M}\overrightarrow{P}-\overrightarrow{k}^{\prime}%
\end{align*}

where the momentum $\overrightarrow{P}=\overrightarrow{p}_{e}+\overrightarrow
{p}_{h}$ is the exciton momentum, i.e. the total momentum of the electron-hole
pair forming the bound state. The wave functions $\psi_{n}\left(
\overrightarrow{r}\right)  $ and $\psi_{\overrightarrow{p}}\left(
\overrightarrow{r}\right)  $\ of the discreet and continuous parts of the
internal exciton spectrum obey the Schredinger equation%

\begin{align*}
\left(  -\frac{\overrightarrow{\nabla}^{2}}{2m^{\ast}}-\frac{1}{\mid
\overrightarrow{r}\mid}\right)  \psi_{n}\left(  \overrightarrow{r}\right)   &
=E_{ex}^{\left(  n\right)  }\psi_{n}\left(  \overrightarrow{r}\right) \\
\left(  -\frac{\overrightarrow{\nabla}^{2}}{2m^{\ast}}-\frac{1}{\mid
\overrightarrow{r}\mid}\right)  \psi_{\overrightarrow{p}}\left(
\overrightarrow{r}\right)   &  =E_{ex}\left(  \overrightarrow{p}\right)
\psi_{\overrightarrow{p}}\left(  \overrightarrow{r}\right)
\end{align*}

The normalized solution of this equation in 2D for the lower energy state can
be written in the momentum representation as%

\[
\psi_{0}\left(  \overrightarrow{p}\right)  =\frac{4\sqrt{\pi}}{\left(
1+\overrightarrow{p}^{2}\right)  ^{3/2}}%
\]

The normalization of this wave function has the form $\int\frac{d^{2}%
p}{\left(  2\pi\right)  ^{2}}\psi_{0}^{2}\left(  \overrightarrow{p}\right)
=1$.

We suppose that the energy $E$ is near the lower bound energy of the exciton
$E_{ex}^{\left(  0\right)  }$. Separating the pole and the non-pole terms we
can write the vertex $\Gamma_{eh}$ as%

\begin{equation}
\Gamma_{eh}\left(  P,\overrightarrow{k},\overrightarrow{k}^{\prime}\right)
=\frac{\left(  E_{ex}^{\left(  0\right)  }-\overrightarrow{k}^{2}/2m^{\ast
}\right)  \left(  E_{ex}^{\left(  0\right)  }-\left(  \overrightarrow
{k}^{\prime}\right)  ^{2}/2m^{\ast}\right)  \psi_{0}\left(  \overrightarrow
{k}\right)  \overline{\psi}_{0}\left(  \overrightarrow{k}^{\prime}\right)
}{\Omega+\mu_{e-h}-\frac{\overrightarrow{P}^{2}}{2M}-E_{ex}^{\left(  0\right)
}+i\delta}+\Gamma_{eh}^{\left(  c\right)  } \label{Ga1}%
\end{equation}
\ \ \ \ 

where the term $\Gamma_{eh}^{\left(  c\right)  }$ is the non-pole part of the
vertex $\Gamma_{eh}$, which can be considered as a constant of the order of
unity $\Gamma_{eh}^{\left(  c\right)  }\sim1$; the collection of the
summarized frequency and the summarized momentum is $P=\left(  \Omega
,\overrightarrow{P}\right)  $. Due to the smallness of the incoming and
outgoing momentums $\mid\overrightarrow{p}_{e,h}\mid,\mid\overrightarrow
{p}_{e,h}^{\prime}\mid<<\hbar/a_{B}=1$ the vertex $\Gamma_{eh}$ can be
represented in the form%

\begin{equation}
\Gamma_{eh}\left(  P\right)  =\frac{F}{\Omega+\mu_{e-h}-\frac{\overrightarrow
{P}^{2}}{2M}-E_{ex}^{\left(  0\right)  }+i\delta}+\Gamma_{eh}^{\left(
c\right)  } \label{Ga2}%
\end{equation}

where the constant $F$ in the case of the small incoming and outgoing
momentums can be represented as%

\[
F=\left(  E_{ex}^{\left(  0\right)  }\psi_{0}\left(  0\right)  \right)  ^{2}%
\]

\ Near the pole of the vertex$\ \Gamma_{eh}$ the term $\Gamma_{eh}^{\left(
c\right)  }\sim1$ can be neglected. The vertex $\Gamma_{eh}$\ can be decoupled
by the introduction of the virtual exciton field $b$, and the action of the
system can be written in the form \ %

\begin{equation}
S_{e,h,ex}^{\left(  s\right)  }=%
{\displaystyle\oint}
d^{3}x\left\{
\begin{array}
[c]{c}%
{\displaystyle\sum\limits_{\alpha=e,h}}
\overline{\psi}_{\alpha}\left(  i\partial_{t}-\xi_{\alpha}\right)
\psi_{\alpha}-\frac{1}{2}\int d^{2}r^{\prime}\left(  \overline{\psi}_{e}%
\psi_{e}-\overline{\psi}_{h}\psi_{h}\right)  \widehat{U}_{C}\left(
\overline{\psi}_{e}\psi_{e}-\overline{\psi}_{h}\psi_{h}\right)  +\\
+\overline{b}\left(  i\partial_{t}-\xi_{ex}\right)  b-\sqrt{F}\left(
\overline{b}\psi_{e}\psi_{h}+\overline{\psi}_{h}\overline{\psi}_{e}b\right)
\end{array}
\right\}  \label{Sehex}%
\end{equation}

where%

\begin{equation}
\xi_{ex}=\frac{\widehat{p}^{2}}{2M}-\mu_{e-h}+E_{ex}^{\left(  0\right)  }
\label{ExcSpec}%
\end{equation}

\[
E_{ex}^{\left(  0\right)  }=-E_{B}%
\]

$E_{B}$ is the bound energy of the internal exciton ground state. As it will
be seen lower, the chemical potential for excitons is connected with
$\mu_{e-h}$ \ as%

\begin{equation}
\mu_{ex}=\mu_{e-h}-E_{ex}^{\left(  0\right)  }-<T_{e,h}^{\left(  0\right)  }>
\label{EexcR}%
\end{equation}

at that, the value $<T_{e,h}^{\left(  0\right)  }>$ can be written as%

\begin{equation}
<T_{e,h}^{\left(  0\right)  }>=G_{e}^{\left(  0\right)  }\left(  x-x^{\prime
}\right)  G_{h}^{\left(  0\right)  }\left(  x-x^{\prime}\right)  =%
{\displaystyle\int\limits_{\mid\overrightarrow{p}\mid>\Lambda}}
\frac{d^{2}p}{\left(  2\pi\right)  ^{2}}\frac{1}{p^{2}/2m^{\ast}} \label{T0}%
\end{equation}

It is convenient to transfer from the Fermi fields of electron-hole plasma to
the density variables, i.e. the density functional with respect to the Fermi
fields. This transition can be accomplished by the introduction of the
functional $\delta$-function to the statistical sum%

\begin{align*}
Z  &  =\int D\psi_{\alpha}D\overline{\psi}_{\alpha}DbD\overline{b}\exp\left(
iS_{e,h,ex}\right)  =\\
&  =\int D\psi_{\alpha}D\overline{\psi}_{\alpha}DbD\overline{b}Dn_{\alpha}%
\exp\left(  iS_{e,h,ex}\right)
{\displaystyle\prod\limits_{\alpha,x}}
\delta\left[  \overline{\psi}_{\alpha}\psi_{\alpha}-n_{\alpha}\right]
\end{align*}

Using the Fourier representation for the functional $\delta$-function%

\[%
{\displaystyle\prod\limits_{x}}
\delta\left[  \overline{\psi}_{\alpha}\psi_{\alpha}-n_{\alpha}\right]  =\int
DV_{\alpha}\exp\left(  -i%
{\displaystyle\oint}
d^{3}xV_{\alpha}\left(  x\right)  \left[  \overline{\psi}_{\alpha}\psi
_{\alpha}-n_{\alpha}\right]  \right)
\]

we can rewrite the action in the form%

\begin{equation}
S_{e,h,ex}\left[  \psi_{\alpha},b,V_{\alpha},n_{\alpha}\right]  =%
{\displaystyle\oint}
d^{3}x\left\{
\begin{array}
[c]{c}%
\left(
\begin{array}
[c]{cc}%
\overline{\psi}_{e}, & \psi_{h}%
\end{array}
\right)  \left(
\begin{array}
[c]{cc}%
i\partial_{t}-\xi_{e}-V_{e} & b\\
\overline{b} & i\partial_{t}-\xi_{h}-V_{h}%
\end{array}
\right)  \left(
\begin{array}
[c]{c}%
\psi_{e}\\
\overline{\psi}_{h}%
\end{array}
\right)  +\\
+%
{\displaystyle\sum\limits_{\alpha}}
V_{\alpha}n_{\alpha}-\frac{1}{2}\left(  n_{e}-n_{h}\right)  \widehat{U}%
_{C}\left(  n_{e}-n_{h}\right)  +\overline{b}\left(  i\partial_{t}-\xi
_{ex}\right)  b
\end{array}
\right\}  \label{Sex0}%
\end{equation}

where%

\begin{equation}
Z=\int D\psi_{\alpha}D\overline{\psi}_{\alpha}DbD\overline{b}Dn_{\alpha
}DV_{\alpha}\exp\left(  iS_{e,h,ex}\left[  \psi_{\alpha},b,V_{\alpha
},n_{\alpha}\right]  \right)  \label{Zex}%
\end{equation}

It is convenient to introduce new variables instead of $n_{\alpha}$ and
$V_{\alpha}$, namely,%

\begin{align}
n_{e}  &  =\frac{n+\delta n}{2};\text{ \ \ \ \ \ \ \ }n_{h}=\frac{n-\delta
n}{2}\label{Vn}\\
V_{e}  &  =V+\delta V;\text{ \ \ \ \ \ \ \ }V_{h}=V-\delta V\nonumber
\end{align}

Note that the fields $\delta n$ and $\delta V$ correspond to the fluctuating
violation of the electro-neutrality in the electron-hole plasma.\ The
integrals over the fields $\psi_{\alpha}$ and $\delta n$ in Eq.(\ref{Zex}%
)\ are Gauss integrals and can be calculated. As a result we obtain the action
in the form%

\begin{align}
S_{ex}\left[  b,V_{\alpha},n_{\alpha}\right]   &  =-iSp\ln\left[  \left(
\begin{array}
[c]{cc}%
i\partial_{t}-\xi_{e}-\left(  V+\delta V\right)  & b\\
\overline{b} & i\partial_{t}-\xi_{h}-\left(  V-\delta V\right)
\end{array}
\right)  \right]  +\label{Sex}\\
&  +%
{\displaystyle\oint}
d^{3}x\left\{  Vn+\frac{1}{2}\delta V\widehat{U}_{C}^{-1}\delta V+\overline
{b}\left(  i\partial_{t}-\xi_{ex}\right)  b\right\} \nonumber
\end{align}

The supposition of the low exciton density gives the possibility to expand in
powers of the field $b$. Integrating over the fields $\delta V$\ and
neglecting the powers higher than $b^{4}$\ we obtain%

\begin{align}
S_{ex}\left[  b,V_{\alpha},n_{\alpha}\right]   &  =-iSp\ln\left[  \left(
\begin{array}
[c]{cc}%
i\partial_{t}-\xi_{e}-V & 0\\
0 & i\partial_{t}-\xi_{h}-V
\end{array}
\right)  \right]  +\label{Sex2}\\
&  +iSp\left\{  \frac{1}{2}\left[  \left(
\begin{array}
[c]{cc}%
0 & b\\
\overline{b} & 0
\end{array}
\right)  \widehat{G}^{\left(  0\right)  }\left[  V\right]  \right]
^{2}\right\}  +iSp\left\{  \frac{1}{4}\left[  \left(
\begin{array}
[c]{cc}%
0 & b\\
\overline{b} & 0
\end{array}
\right)  \widehat{G}^{\left(  0\right)  }\left[  V\right]  \right]
^{4}\right\}  -\nonumber\\
&  -E_{corr}\left[  V\right]  +%
{\displaystyle\oint}
d^{3}x\left\{  Vn+\overline{b}\left[  i\partial_{t}-\xi_{ex}\right]  b\right\}
\nonumber
\end{align}

The functional $E_{corr}\left[  V\right]  $ is the correlation energy. This
functional is represented as the sum of all simply connected closed diagrams
dressed by the internal Coulomb interaction lines and the lines of the
external field V. In the case of constant V the correlation energy
$E_{corr}\left[  V\right]  $ can be represented in the well-known form%

\begin{equation}
E_{corr}\left[  V\right]  ==-i\frac{1}{2}Sp\ln\left[  1-\Pi\left[  V\right]
\widehat{U}_{C}\right]  =i\frac{1}{2}%
{\displaystyle\int\limits_{0}^{1}}
dg\int\frac{d\omega d^{2}k}{\left(  2\pi\right)  ^{3}}\frac{U_{C}\left(
k\right)  \Pi\left[  V\right]  }{1-gU_{C}\left(  k\right)  \Pi\left[
V\right]  }\label{Ecorr}%
\end{equation}

where $\Pi\left[  V\right]  $ is the total polarization operator. The simple
transformations of the second and third terms in (\ref{Sex2})\ give the action
$S_{ex}$ in the form%

\begin{align}
S_{ex}\left[  V,n\right]   &  =-iSp\ln\left[  \left(
\begin{array}
[c]{cc}%
i\partial_{t}-\xi_{e}-V & 0\\
0 & i\partial_{t}-\xi_{h}-V
\end{array}
\right)  \right]  -E_{corr}\left[  V\right]  +\label{Sex3}\\
&  +%
{\displaystyle\oint}
d^{3}x\left\{  Vn+\overline{b}\left[  i\partial_{t}-\xi_{ex}\right]
b-\frac{1}{2}\gamma_{ex}\left(  \overline{b}b\right)  ^{2}-\gamma
_{ex-e,h}\left(  \overline{b}b\right)  n\right\} \nonumber
\end{align}

The value $<T_{e,h}^{\left(  0\right)  }>$ has the form (\ref{T0}). The
coupling constant $\gamma_{ex}$ is the exciton-exciton scattering amplitude;
and $\gamma_{ex-e,h}=\gamma_{ex-e}+\gamma_{ex-h}$ where $\gamma_{ex-e}$,
$\gamma_{ex-h}$ are the exciton-electron and exciton-hole scattering
amplitudes, respectively.

The first order expansion of the first term in Eq.(\ref{Sex3}) over the field
V gives the change $Vn\rightarrow V\left(  n-<n>\right)  $, where $<n>$ is the
average density of the electron-hole plasma in stationary state $<n>=\frac
{\int d^{2}rn\left(  t,\overrightarrow{r}\right)  }{Vol}$, where $Vol$ is the
volume of the system. The second order expansion of the first term in
Eq.(\ref{Sex3}) in series of the field $V$ and the integration over this field gives%

\begin{equation}
S_{ex}\left[  b,n\right]  =%
{\displaystyle\oint}
d^{3}x\left\{
\begin{array}
[c]{c}%
\left[  \frac{1}{2}\left(  n-<n>\right)  \left(  \Pi^{\left(  0\right)
}\right)  ^{-1}\left(  n-<n>\right)  -E_{corr}\left[  n\right]  \right]  +\\
+\overline{b}\left[  i\partial_{t}+\mu_{ex}-\mu_{e-h}-\xi_{ex}\right]
b-\frac{1}{2}\gamma_{ex}\left(  \overline{b}b\right)  ^{2}-\gamma
_{ex-e,h}\left(  \overline{b}b\right)  n
\end{array}
\right\}  \label{Seff}%
\end{equation}

where $\Pi^{\left(  0\right)  }$ is the polarization operator of the form
$\Pi^{\left(  0\right)  }\left(  x-y\right)  =-i%
{\displaystyle\sum\limits_{\alpha}}
G_{\alpha}^{\left(  0\right)  }\left(  x-y\right)  G_{\alpha}^{\left(
0\right)  }\left(  y-x\right)  $. Expending in series of the field V we
suppose the smallness of this field. This supposition is correct when the
fluctuations of the density n are small compared with $<n>$. This supposition
is correct for the beginning of the density fluctuations growth. The
derivation of the effective action in the supposition of the smooth variation
of the density n at the scales lager than the average distance between
particles without supposition of the smallness of the density fluctuations
will be considered in the next paper.

\section{Instability of the system and the density modulation of the
electron-hole plasma}

Using the effective action for the interacting electron-hole plasma and the
exciton gas $S_{ex}\left[  b,n\right]  $ Eq.(\ref{Seff}) we show that the
homogeneous state of this system is unstable, and due to the Coulomb
correlations the low density electron-hole plasma should form the
inhomogeneous state. The formation of the inhomogeneous state of the low
density electron-hole plasma is the result of the tendency of the low density
plasma to collapse to the liquid electron-hole drops. The interaction of this
plasma with the exciton gas results in the existence of the self-consistent
nonhomogeneous periodic field acting on the exciton system as the periodic
external field. This periodic field results in the density modulation of the
exciton system.

To analyze the effective action for simplicity we put $m=m_{e}=m_{h}=1$. The
polarization operator $\Pi^{\left(  0\right)  }\left(  \omega,\overrightarrow
{k}\right)  $ for $\mid\overrightarrow{k}\mid<<p_{F}$\ is calculated as%

\begin{equation}
\Pi^{\left(  0\right)  }\left(  \omega,\overrightarrow{k}\right)  =\left\{
\begin{array}
[c]{c}%
\frac{m}{2\pi}\frac{\left(  kV_{F}\right)  ^{2}}{\omega^{2}}\text{
\ \ \ \ \ \ \ \ \ \ \ \ \ \ \ \ \ \ \ \ \ \ for }\omega>>kV_{F}\\
-\frac{1}{\pi}m\left(  1+i\pi\left(  \frac{\omega}{kV_{F}}\right)  \right)
\text{ \ \ for }\omega<<kV_{F}%
\end{array}
\right\}  \label{Pai}%
\end{equation}

The correlation energy $E_{corr}\left[  n\right]  $ for small densities
$na_{B}^{2}<<1$ can be calculated as%

\begin{equation}
E_{corr}\left[  n\right]  =-An^{4/3}\label{Ecorr1}%
\end{equation}

where the constant A is estimated as $A\sim1$. The main contribution to the
integral for $E_{corr}\left[  n\right]  $ (\ref{Ecorr}) is given by the
momentums $k_{0}\sim n^{1/3}>>p_{F}\sim n^{1/2}$ and the frequencies
$\omega_{0}\sim k_{0}^{2}$. Note that the Hartree-Fock energy for this system
is estimated as%

\begin{equation}
E_{H-F}\left[  n\right]  \sim-\frac{n^{2}}{p_{F}}\sim-n^{3/2}\label{Ehf}%
\end{equation}

Thus, the contribution of the Hartree-Fock energy can be neglected compared
with the correlation energy $E_{H-F}\left(  n\right)  <<E_{corr}\left(
n\right)  $ for $na_{B}^{2}<<1$.

The substitution of the polarization operator $\Pi^{\left(  0\right)  }\left(
\omega,\overrightarrow{k}\right)  $ Eq.(\ref{Pai}) for $\omega>>kV_{F}$\ to
Eq.(\ref{Seff}) gives%

\[
S_{ex}\left[  n\right]  =%
{\displaystyle\oint}
d^{3}x\left\{
\begin{array}
[c]{c}%
\left[  \left(  \Delta n\right)  \left[  \frac{1}{2\pi}\frac{\omega^{2}%
}{\left(  kV_{F}\right)  ^{2}}\right]  \left(  \Delta n\right)  -E_{corr}%
\left(  n\right)  \right]  +\\
+\overline{b}\left[  i\partial_{t}-\xi_{ex}\right]  b-\frac{1}{2}\gamma
_{ex}\left(  \overline{b}b\right)  ^{2}-\gamma_{ex-e,h}\left(  \overline
{b}b\right)  n
\end{array}
\right\}
\]

where $\Delta n=n-<n>\ $\ and expanding in $\Delta n$\ the functional
$E_{corr}\left[  n\right]  $\ we obtain%

\begin{equation}
S_{ex}\left[  \delta n\right]  =%
{\displaystyle\oint}
d^{3}x\left\{
\begin{array}
[c]{c}%
\left(  \Delta n\right)  \left[  \frac{1}{2\pi}\frac{\omega^{2}}{\left(
kV_{F}\right)  ^{2}}-\frac{1}{2}E_{corr}^{\prime\prime}\left(  <n>\right)
\right]  \left(  \Delta n\right)  +\overline{b}\left[  i\partial_{t}-\xi
_{ex}\right]  b-\\
-\frac{1}{2}\gamma_{ex}\left(  \overline{b}b\right)  ^{2}-\gamma
_{ex-e,h}\left(  \overline{b}b\right)  <n>-\gamma_{ex-e,h}\left(  \overline
{b}b\right)  \Delta n
\end{array}
\right\}  \label{Sn1}%
\end{equation}

In these expressions the frequency $\omega$ should be considered as
$\omega=i\partial_{t}$. The term linear over $\Delta n$, namely,
$E_{corr}^{\prime}\left(  <n>\right)  \Delta n$, can be omitted if the total
number of particles is fixed, i.e., $\int\Delta n=0$. The substitution of
$E_{corr}^{\prime\prime}\left(  <n>\right)  $ (\ref{Ecorr1}) to (\ref{Sn1})\ gives%

\begin{equation}
S_{ex}\left[  \delta n\right]  =%
{\displaystyle\oint}
d^{3}x\left\{
\begin{array}
[c]{c}%
\Delta n\left[  \frac{1}{2\pi}\frac{\omega^{2}}{\left(  kV_{F}\right)  ^{2}%
}+\frac{2}{9}A<n>^{-2/3}\right]  \Delta n+\overline{b}\left[  i\partial
_{t}-\xi_{ex}\right]  b+\\
-\frac{1}{2}\gamma_{ex}\left(  \overline{b}b\right)  ^{2}-\gamma
_{ex-e,h}\left(  \overline{b}b\right)  <n>-\gamma_{ex-e,h}\left(  \overline
{b}b\right)  \Delta n
\end{array}
\right\}  \label{Sn2}%
\end{equation}

From the first term in Eq. (\ref{Sn2}) it can be seen that the system is
unstable relative to the phonon oscillations of the electron-hole plasma
density. The sound velocity of the density oscillations has the imaginary
value $\mid c\mid^{2}=\left(  \frac{4\pi}{9}A<n>^{-2/3}V_{F}^{2}\right)  $ .
Note that the sound character of the considering density fluctuations is not
surprising, since these fluctuations do not violate the electro-neutrality of
the electron-hole plasma. This instability results in the growth of the
density fluctuations of the plasma with the increment of the growth%

\[
\zeta=\left(  \frac{4\pi}{9}A\right)  ^{1/2}<n>^{-1/3}kV_{F}\sim
k<n>^{1/6}>>kV_{F}%
\]

This increment of the growth corresponds to the frequency which is much larger
than $kV_{F}$, $\zeta>>kV_{F}$. Due to this inequality we can use the
asymptotic of the polarization operator $\Pi^{\left(  0\right)  }\left(
\omega,k\right)  \sim\frac{\left(  kV_{F}\right)  ^{2}}{\omega^{2}}$ being
correct for the large frequencies $\omega>>kV_{F}$.

As the result, the stationary state of the electron-hole plasma should be
nonhomogeneous, besides, the modulation of the density has the periodic character.

The modulated density of the electron-hole plasma results in the existence of
the self-consistent field acting on the exciton system. This field has the
form (\ref{Sn2})%

\[
V\left(  \overrightarrow{r}\right)  =\gamma_{ex-e,h}\Delta n\left(
\overrightarrow{r}\right)
\]

Note that this external for the exciton system potential plays the essential
role, if the amplitude of this potential is larger than the chemical potential
of the exciton gas. This condition can be obeyed for the sufficiently large
density of the electron-hole plasma, i.e., for the sufficiently large pump.

At the conclusion we estimate the radius of the character non-homogeneity of
the electron-hole plasma from the kinetic consideration \cite{KKin}. We
suppose that the kinetics of the non-homogeneity formation is analogous to the
kinetics of the electron-hole liquid drop formation or the phase immiscibility
and can be described by the equation%

\begin{equation}
\partial_{t}\left(  \pi R^{2}n_{0}\right)  =2\pi RV_{F}n+2\pi RV_{ex}%
n_{ex}-\pi R^{2}\frac{n_{0}}{\tau} \label{KinDrop}%
\end{equation}

The first term in the right hand side of the equation (\ref{KinDrop}) is the
incoming term and the second one is the outgoing term; $V_{F}n$\ is the
current of Fermi particles incoming into the drop of the radius R;
$V_{ex}n_{ex}$\ is the current of excitons incoming into the drop, $V_{ex}%
$\ is the average velocity of excitons which can be written as $V_{ex}%
\sim\frac{\sqrt{\gamma_{ex}n_{ex}}}{m}$; n is the density of the electron-hole
plasma in the gas phase, $n_{ex}$ is the density of excitons, and $n_{0}$ is
the density of the electron-hole liquid within the drop; $\tau$ is the
lifetime of the carriers. There are no vapor terms in Eq.(\ref{KinDrop}) since
T=0, and particles can not overcome work function from the liquid drop. The
equilibrium value of the radius R can be found from Eq.(\ref{KinDrop}) as%

\begin{equation}
R_{0}=2\frac{V_{F}n+V_{ex}n_{ex}}{n_{0}}\tau\label{R0}%
\end{equation}

Note that the radiuses larger than $R_{0}$ are not profitable due to the
different dependence on R of the incoming and outgoing terms - the first one
is the surface term and the second one has the volume character. From the
point of view of the phase immiscibility the liquid phase is the electron-hole
liquid drops and the gas phase is the mixture of the exciton gas and the
electron-hole plasma of the small density.

Finally, the set of drops forms the periodic structure. We suppose that this
is the consequence of the homogeneity of the system in average. This periodic
structure forms the external periodic field for the exciton gas. If this Bose
gas could be considered as the ideal, it should be localized at the minima of
the periodic external potential.

\end{document}